\documentclass[aps,prc,reprint,nofootinbib]{revtex4-1}
\usepackage{graphicx}   
\usepackage{latexsym}   
\usepackage{enumerate}
\usepackage{rotating,booktabs,multirow}
\begin{document}
\title{Delta mass shift as a thermometer of kinetic decoupling in Au+Au reactions at 1.23~AGeV}
\author{Tom Reichert$^{1,2}$, Paula~Hillmann$^{1,2,3,4}$, Ayut Limphirat$^{5}$, Christoph Herold$^{5}$,  Marcus~Bleicher$^{1,2,3,4}$}

\affiliation{$^1$ Institut f\"ur Theoretische Physik, Goethe Universit\"at Frankfurt, Max-von-Laue-Strasse 1, D-60438 Frankfurt am Main, Germany}
\affiliation{$^2$ Frankfurt Institute for Advanced Studies, Ruth-Moufang-Str. 1, 60438 Frankfurt am Main, Germany}
\affiliation{$^3$ GSI Helmholtzzentrum f\"ur Schwerionenforschung GmbH, Planckstr. 1, 64291 Darmstadt , Germany}
\affiliation{$^4$ John von Neumann-Institut f\"ur Computing, Forschungzentrum J\"ulich,
52425 J\"ulich, Germany}
\affiliation{$^5$ School of Physics and Center of Excellence in High Energy Physics $\&$ Astrophysics, Suranaree University of Technology, Nakhon Ratchasima 30000, Thailand}

\begin{abstract}
The HADES experiment at GSI will soon provide data on the production and properties of $\Delta$(1232) baryons from Au+Au reactions at 1.23 AGeV. Using the UrQMD model, we predict the yield and spectra of $\Delta$(1232) resonances. In addition we show that one expects to observe a mass shift of the $\Delta$(1232) resonance on the order of 50 MeV in the reconstructable $\Delta$(1232) mass distribution. This mass shift can be understood in terms of late stage $\Delta$(1232) formation with limited kinetic energy. We show how the mass shift can be used to constrain the kinetic decoupling temperature of the system.
\end{abstract}

\maketitle
\section{Introduction}

The fundamental theory of strong interaction, called Quantum Chromodynamics (QCD), can be tested with matter under extreme conditions of density and temperature. These conditions are found in nature in the interior of compact stellar objects, e.g. neutron stars, the accretion discs of black holes or billions of years ago just micro seconds after the big bang. Nowadays such conditions can be recreated in the collision of massive atomic nuclei in today's largest particle accelerators. A prime question of the field is the exploration of the phase structure of QCD, i.e. the exploration of the active degrees of freedom and their properties in the whole density-temperature plane. At very high collision energies, e.g. top RHIC energy and LHC energies, the exploration of the deconfined state of QCD, the Quark Gluon Plasma is in the center of attention. At  lower collision energies the reachable energy densities and temperatures decrease, however the net-baryon density increases, opening the road to explore its effects. Modifications of the spectral function due to temperature and density are summarized as in-medium effects \cite{Aarts:2018glk,Aarts:2017rrl}. Such in-medium effects have been extensively  studied at RHIC \cite{Torrieri:2001ue,Bleicher:2002dm,Broniowski:2003ax,Kolb:2003bk,Geurts:2012rv,Makek:2016rnn} energies and SPS \cite{Arnaldi:2006jq,Damjanovic:2007qm,vanHees:2007th,Endres:2014zua}, for the latest results from the LHC we refer to \cite{Albuquerque:2018kyy,ALICE:2018ewo,Tripathy:2018ehz}. Traditionally, the $\rho$ meson and its decay into di-leptons was the most studied resonance. These studies where supplemented by the exploration of the decay channel $\rho\rightarrow\pi+\pi$ which selects mainly the final stage $\rho$-mesons \cite{Bleicher:2002dm,Pratt:2003vb,Markert:2005ms,Fachini:2008zz,Knospe:2015nva} as it will be discussed below. While HADES has also contributed strongly to the exploration of in-medium effects using di-leptons, a direct measurement of the hadronic decay of the $\rho$-meson is still missing from this experiment. However, HADES has recently reported data on reconstructed $\Delta$(1232) resonances in the $\pi+N$ channel \cite{Kornakov:2019talk}). Most notably are these results as they indicate a substantial mass shift of the $\Delta$(1232) in Au+Au reactions at 1.23 AGeV beam energy. In this paper we want to investigate these findings and will interpret them as a late stage regeneration effect. We also suggest that the low transverse momentum mass shift might allow to extract the temperature of the resonance decoupling stage.

\subsection{The UrQMD model and the reconstruction of resonances\label{reco}}
For the present investigation we use the Ultra relativistic Quantum Molecular Dynamics (UrQMD) transport model in version 3.4 \cite{Bass:1998ca,Bleicher:1999xi}. This model has a well established history for the description of hadron yields and spectra over a broad range of energies. UrQMD is based on the covariant propagation of hadrons and their interactions by potentials and/or inelastic cross sections. It includes a large range of baryonic and mesonic states with resonance masses up to more than 2 GeV. The model has been extensively tested in the GSI energy regime at 1.23~AGeV in Au+Au collisions, at 1.76~AGeV in Ca+Ca collisions and at 1.76~AGeV in Ar+KCl collisions under investigation in this work, see e.g. \cite{Steinheimer:2015sha,Endres:2015fna,Steinheimer:2016vzu,Hillmann:2018nmd}.  

During the time evolution of the system, baryon resonances are formed either by the scattering of two baryons, e.g. $pp\rightarrow \Delta^++p$, or by s-channel process e.g. $\pi^++p\rightarrow \Delta^{++}$. If in vacuum, the resonance state propagates until it decays, where the decay time is given by the total width of the resonance. However, the resonance may also acquire an in-medium collisional width, resulting in a shorter life time, due to its interaction with surrounding hadrons. Experimentally a decayed resonance is reconstructed from the invariant mass distribution of its decay daughters. Here one takes all pairs from each single event and subtracts pairs from mixed events to obtain the signal. Of course, exactly the same method can be applied for a transport simulation. However, such an approach has  a major disadvantage as it needs substantial statistics and does not allow to pin down the properties of single resonances in the simulations. Instead of the mixed-event approach, we follow the decay products (daughter particles) of each individual resonance after its decay. The resonance can only be reconstructed, if the momenta of the daughter particles during their evolution through the system are not strongly changed, i.e. that the invariant mass of the combined daughter particles remains within the Delta mass region. To explore this in detail, we investigate four scenarios and follow each particle until kinetic freeze-out  \cite{Johnson:1999fv,Bleicher:2002dm}:
\begin{enumerate}
\item Both daughter particles of the Delta (i.e. pion and nucleon) do not rescatter on their way out. In this case the resonance can be reconstructed in the invariant mass  spectrum. 
\item One of the daughter particles rescatters elastically (probably many times) on the way out, while the other particles leaves the system undisturbed. In this case the invariant mass of the daughter particles is calculated and taken into account, if it within the Delta mass region.
\item Both daughter particles rescatter elastically (probably many times) on the way out. Also for this case we calculate the invariant mass of the daughter particles. However, as the analysis will show, in this case the invariant mass distribution is flat and the Delta can not be reconstructed.
\item One or both daughter particles rescatter inelastically. In this case the information about the invariant mass of the decaying resonance is also lost and the Delta can not be reconstructed.
\end{enumerate}

\subsection{Rapidity and transverse momentum spectra}
\begin{figure}[t]	
\includegraphics[width=1.6\textwidth]{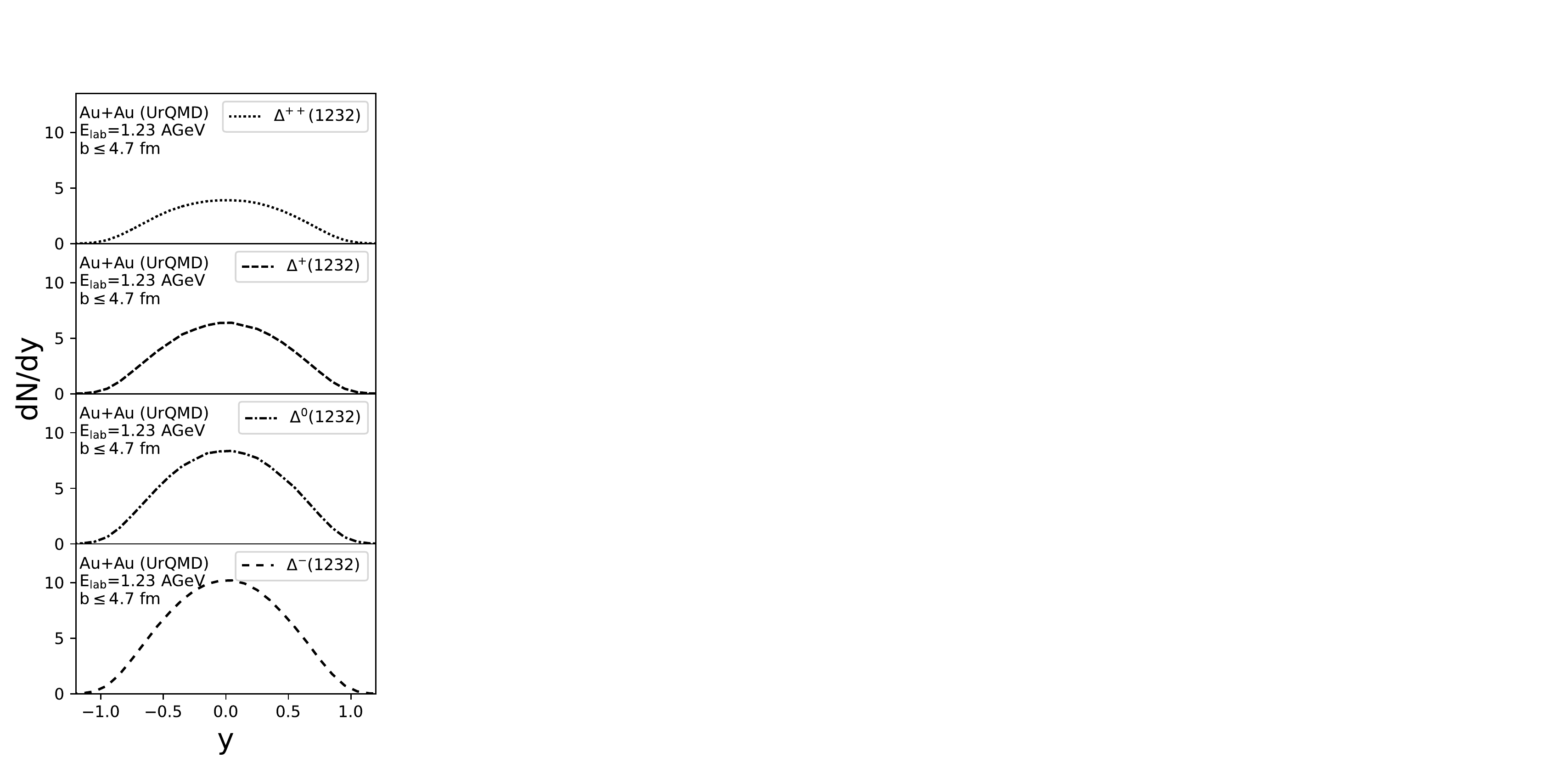}
\caption{[Color online] Rapidity distributions of $\Delta^{++}$(1232), $\Delta^{+}$(1232), $\Delta^{0}$(1232), $\Delta^{-}$(1232) (from top to bottom), summed over all decay channels are shown by lines.
}\label{dndy}
\end{figure}

To begin our exploration, we calculate Au+Au reactions at various centralities for a beam energy of 1.23 AGeV. 
Fig. \ref{dndy} shows the rapidity distributions of $\Delta$(1232) resonances for the most central collisions. The yield is obtained by summing the invariant mass spectrum up to 1.4~GeV/c$^2$. From top to bottom we show $\Delta^{++}$(1232), $\Delta^{+}$(1232), $\Delta^{0}$(1232) and $\Delta^{-}$(1232). All possible decay channels are added up%
\footnote{I.e. we show reconstructable resonances in the channels $\Delta^{++}\rightarrow p+\pi^+$, $\Delta^{+}\rightarrow p+\pi^0$, $\Delta^{+}\rightarrow n+\pi^+$, $\Delta^{0}\rightarrow p+\pi^-$, $\Delta^{0}\rightarrow n+\pi^0$, $\Delta^{-}\rightarrow n+\pi^-$. In the case of the $\Delta^+$(1232) and the $\Delta^0$(1232) one may apply the corresponding Clebsch-Gordon coefficents (shown in Table \ref{table:decay-channels}), if only one outgoing state can be detected.}%
. The model calculations are shown by lines.

\begin{table} [h]
	\centering
	\begin{tabular}{c|c}
		Decay channel of $\Delta$(1232)& Branching ratio\\ 
		\hline 
		\hline
		$\Delta^{++}\rightarrow\pi^{+}+p$& 1\\ 
		\hline 
		$\Delta^{+}\rightarrow\pi^{0}+p$& 2/3\\ 
		\hline
		$\Delta^{+}\rightarrow\pi^{+}+n$&1/3 \\ 
		\hline 
		$\Delta^{0}\rightarrow\pi^{-}+p$&1/3\\ 
		\hline  
		$\Delta^{0}\rightarrow\pi^{0}+n$&2/3\\ 
		\hline 
		$\Delta^{-}\rightarrow\pi^{-}+n$&1\\ 
		\hline 
	\end{tabular} 
	\caption{The possible decay channels of the several charged $\Delta$(1232) resonances and their individual branching ratios calculated from the Clebsch-Gordon coefficients (column 2) are shown.}\label{table:decay-channels}
\end{table}

The first observation is an isospin dependence of the $\Delta$(1232) yield. In line with the strong isospin asymmetry of the gold nucleus we observe an enhancement of the more negatively charged states as compared to the positively charged states.  We focus our comparison on the $\Delta^{++}$(1232) state because it can be unambiguously reconstructed in the $p+\pi^+$ channel. It should also be noted that at such low energies, the $\Delta$(1232) yield is tightly connected to the pion production which is in line with the experimental measurements. 

\begin{figure}[t]	
\includegraphics[width=0.5\textwidth]{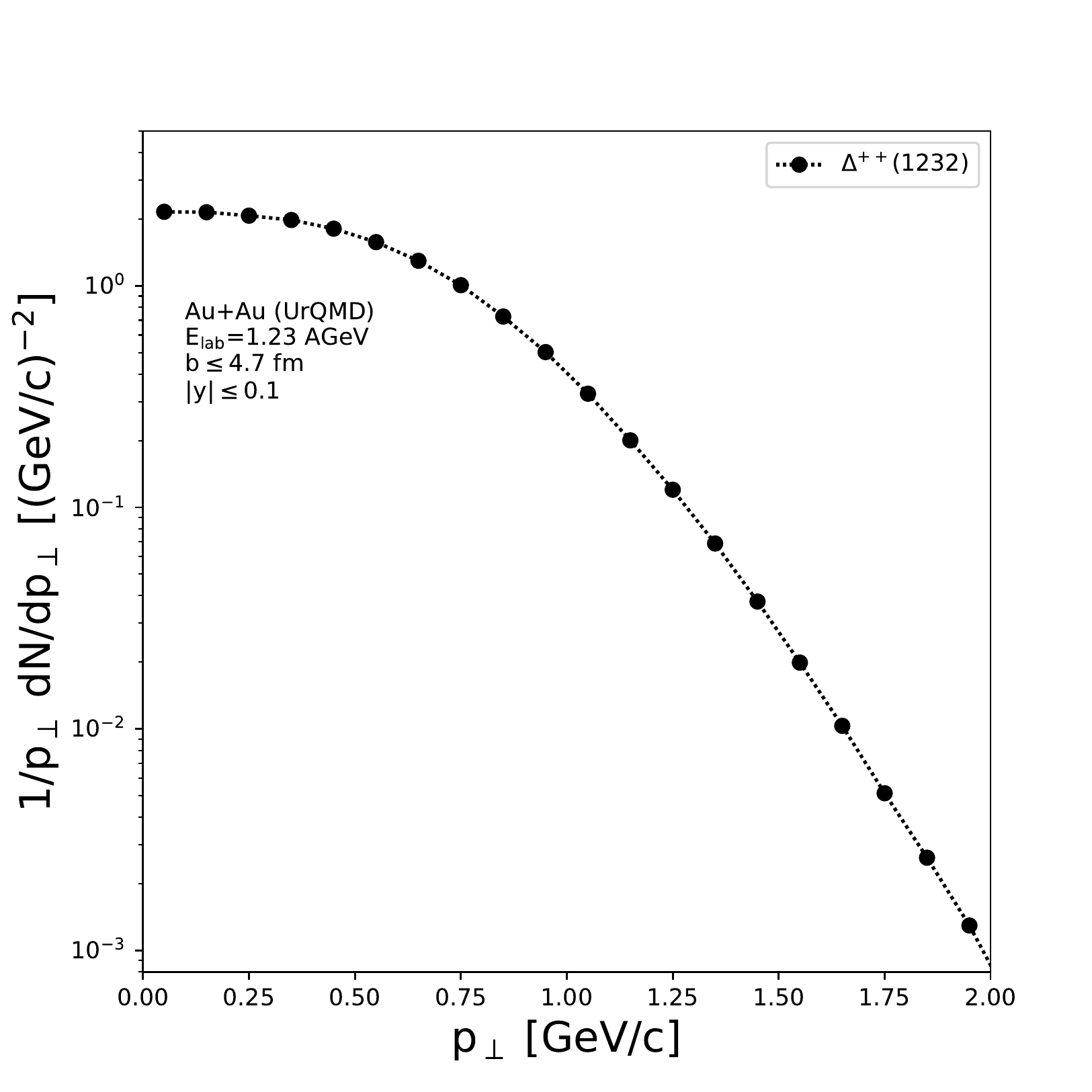}
\caption{[Color online] Transverse momentum distributions at mid-rapidity of reconstructable $\Delta^{++}$(1232) (line with full circles) and of all decaying $\Delta^{++}$(1232) (line with full squares) 
in Au+Au reaction at 1.23 AGeV from UrQMD.}\label{dndpt}
\end{figure}

Next we explore the transverse momentum distribution of the $\Delta$(1232) baryons for central reactions at mid-rapidity in Fig. \ref{dndpt}. The yield is again obtained by summing the invariant mass spectrum up to 1.4~GeV/c$^2$.

\subsection{Mass distribution}
Only resonances decaying near the kinetic freeze-out surface have a substantial chance of being reconstructed in hadronic decay channels in the experiment. Especially short lived resonances like the $\Delta$(1232) decay quickly and are regenerated many times before their final observable decay happens. It is clear that the cross section in the reaction $\Delta\leftrightarrow \pi+N$ and the resulting mass distribution of the $\Delta$(1232) are tightly connected to the kinetic energies of the pions and nucleons in the local rest frame of the matter, which is nothing else than the temperature. Therefore one expects that the mass distribution of the $\Delta$(1232) baryons reconstructed in the $\pi+p$ channel reflects the temperature near the kinetic freeze-out surface of the system.

What is our expectation in this setting? Delta resonances are formed by collisions of a pion and a nucleon. The kinetic energies of the pion and the nucleon (and therefore their center-of-mass energy) depends on the temperature of the system at the time of the interaction. If the temperature is low the center-of-mass energy is low and the Delta resonance is populated at the lower end of the mass distribution. In contrast, if the temperature is high, the Delta is populated on the high mass tail of the distribution. To a first approximation and in the case that the system can be assumed to be described by a temperature, we expect to observe a mass distribution for the $\Delta$(1232) following a standard Breit-Wigner (BW) spectral function multiplied by the thermal weight of each resonance mass state  (often called phase space factor (PS))  \cite{Ilner:2017tab,Fachini:2006up}.  I.e. 
\begin{equation}\label{BWxPS}
P(m_\Delta,p_{\perp}) \propto BW(m_\Delta,m_\Delta^0,\Gamma_\Delta^0) \times PS(m_\Delta,p_{\perp},T)
\end{equation}
with 
\begin{equation}
	BW(m_\Delta,m_\Delta^0,\Gamma_\Delta^0)\propto\frac{\Gamma m_{\Delta}}{\left(m_{\Delta}^2-m_{\Delta}^0\right)^2+(\Gamma m_{\Delta})^2}
\end{equation} 
and 
\begin{equation}
	PS(m_\Delta,p_{\perp},T)\propto\frac{ m_{\Delta}}{\sqrt{m_\Delta^2+p_{\perp}^2}}\exp\left(-\frac{\sqrt{m_\Delta^2+p_{\perp}^2}}{T}\right).
\end{equation}
Here $m_\Delta$ is the mass of the $\Delta$(1232), $m_\Delta^0=1.232$ GeV/c$^2$ is the pole mass of the $\Delta$(1232), $\Gamma_\Delta^0=0.118$ GeV/c$^2$ is the width of the $\Delta$(1232), $p_{\perp}$ is the transverse momentum and $T$ is the temperature. This leads us to two direct observations in line with our discussion above: 1) The average mass (and the most probable mass) of the $\Delta$(1232) decreases with decreasing temperature and 2) the average mass (and the most probable mass) of the $\Delta$(1232) increases with increasing transverse momentum. One should note that this mass shift behaves opposite to the usual in-medium effects that predict a stronger shift with increasing temperature \cite{vanHees:2004vt}. Let us now explore, if this qualitative assumptions and interpretations are supported by a full transport model calculation. Here we employ of course mass dependent decay widths for the resonance and do not use the notion of temperature, but the evolution of the momentum distributions.

\begin{figure}[t]	
\includegraphics[width=0.5\textwidth]{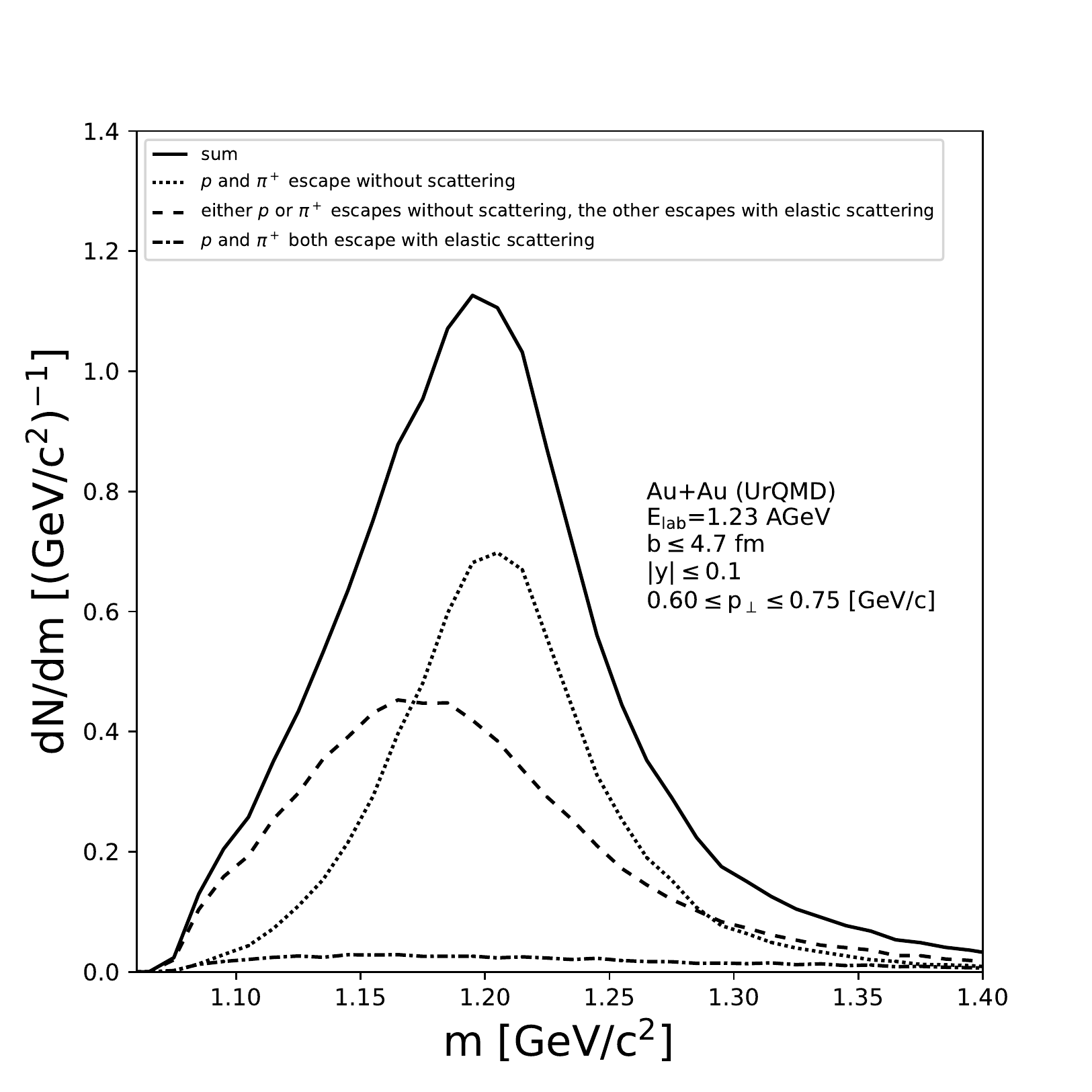}
\caption{[Color online] Invariant mass distribution of reconstructable $\Delta^{++}$(1232) resonances at midrapidity for a given transverse momentum bin in central Au+Au reactions at 1.23 AGeV from UrQMD. The dotted line shows the contribution from I, the dashed line shows the contribution from II the dashed dotted line shows the contribution from III and the full line shows the sum of the different contributions.
}\label{mass_dist}
\end{figure}

Before we explore the mass distributions as a function of transverse momentum and rapidity, let us investigate the influence of the different daughter interactions to reconstruct the Delta mass spectrum in more detail. The different channels for the invariant mass reconstruction were defined in Sec. \ref{reco}. In Fig. \ref{mass_dist} we show the invariant mass distribution of $m_{p+\pi^+}$ for the three reconstruction channels: I) Both daughter particles escape undisturbed, II) one daughter particle rescatters elastically, and III) both daughter particles rescatter elastically. In the first two cases, one obtains contributions within the Delta mass range, showing that also elastically rescattered daughter particles contribute to the Delta yield. In the case where both daughter particles rescatter, no resonance can be reconstructed, because the invariant mass distribution has become flat. It is interesting to note that the elastic rescattering populates the low invariant mass region more strongly, and does therefore also contribute to the apparent mass shift of the Delta resonance.

\begin{figure}[t]	
\includegraphics[width=1.5\textwidth]{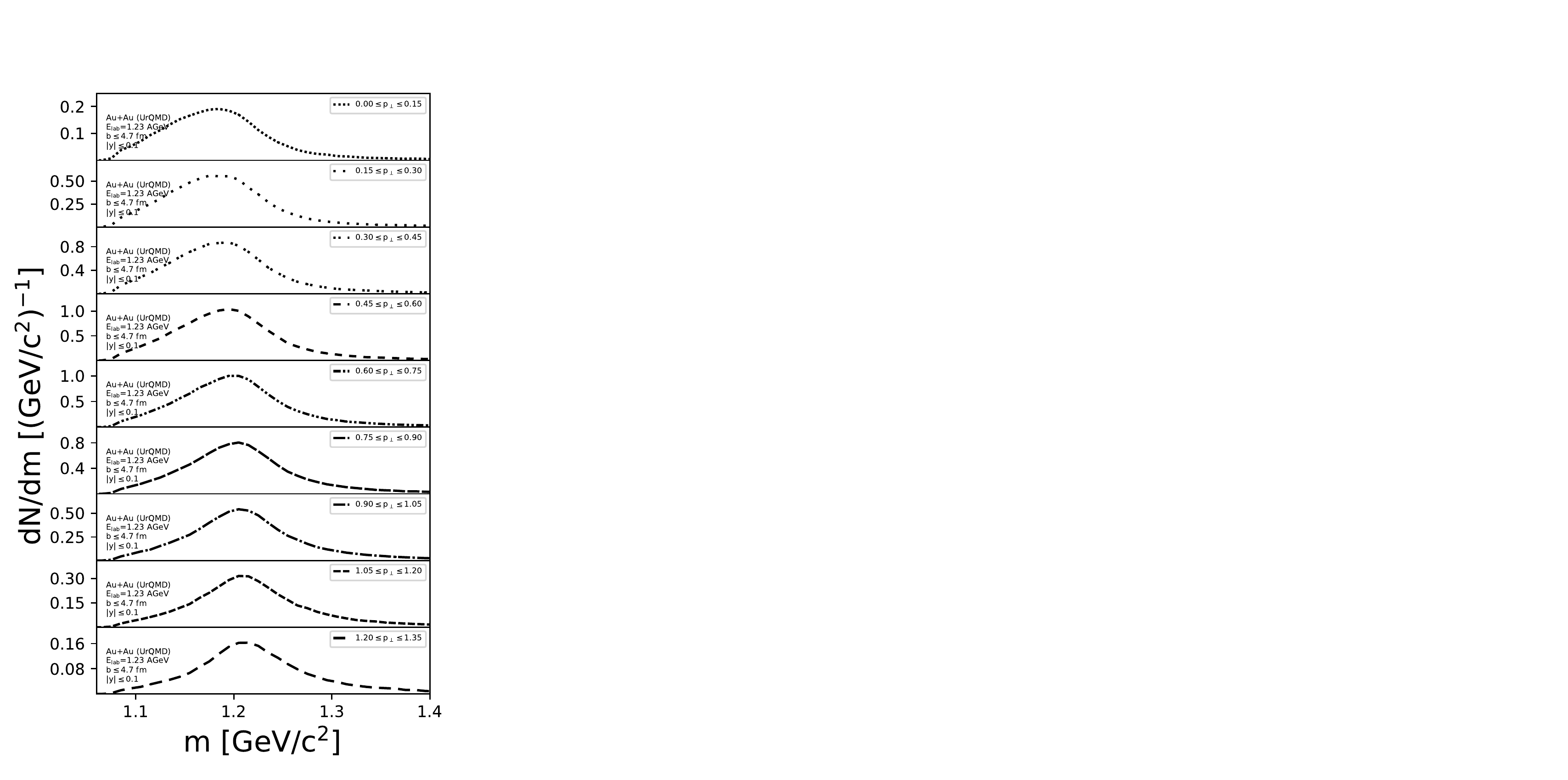}
\caption{[Color online] Mass distribution of $\Delta^{++}$(1232) resonances at midrapidity for various momentum bins ranging from $p_{\perp}=0-0.15$ GeV to $p_{\perp}=1.2-1.35$ GeV (from top to bottom) in central Au+Au reactions at 1.23 AGeV from UrQMD. 
}\label{dndm_pt}
\end{figure}

In Fig. \ref{dndm_pt} we show the mass distribution of reconstructable $\Delta$(1232) baryons obtained from the and the invariant mass distribution of $p+\pi^+$ for several transverse momentum bins in central Au+Au reactions. In line with our discussion above we observe an essentially unmodified $\Delta$(1232) mass distribution at high transverse momenta, while one clearly observes a shift towards lower masses when going from high transverse momenta to low transverse momenta. One should note that the mass shift is limited by the observation method by $\langle m_\Delta\rangle\ge m_p+m_{\pi}$ (due to the reconstruction in the pion+nucleon channel). Thus, this kinetic mass shift leads to a trivial narrowing of the mass distribution towards low transverse momenta that may overshadow a potentially present broadening due to (conventional) in-medium effects.

\subsection{Rapidity and transverse momentum dependence of the $\Delta$ mass shift}
\begin{figure}[t]	
	\includegraphics[width=
	0.5\textwidth]{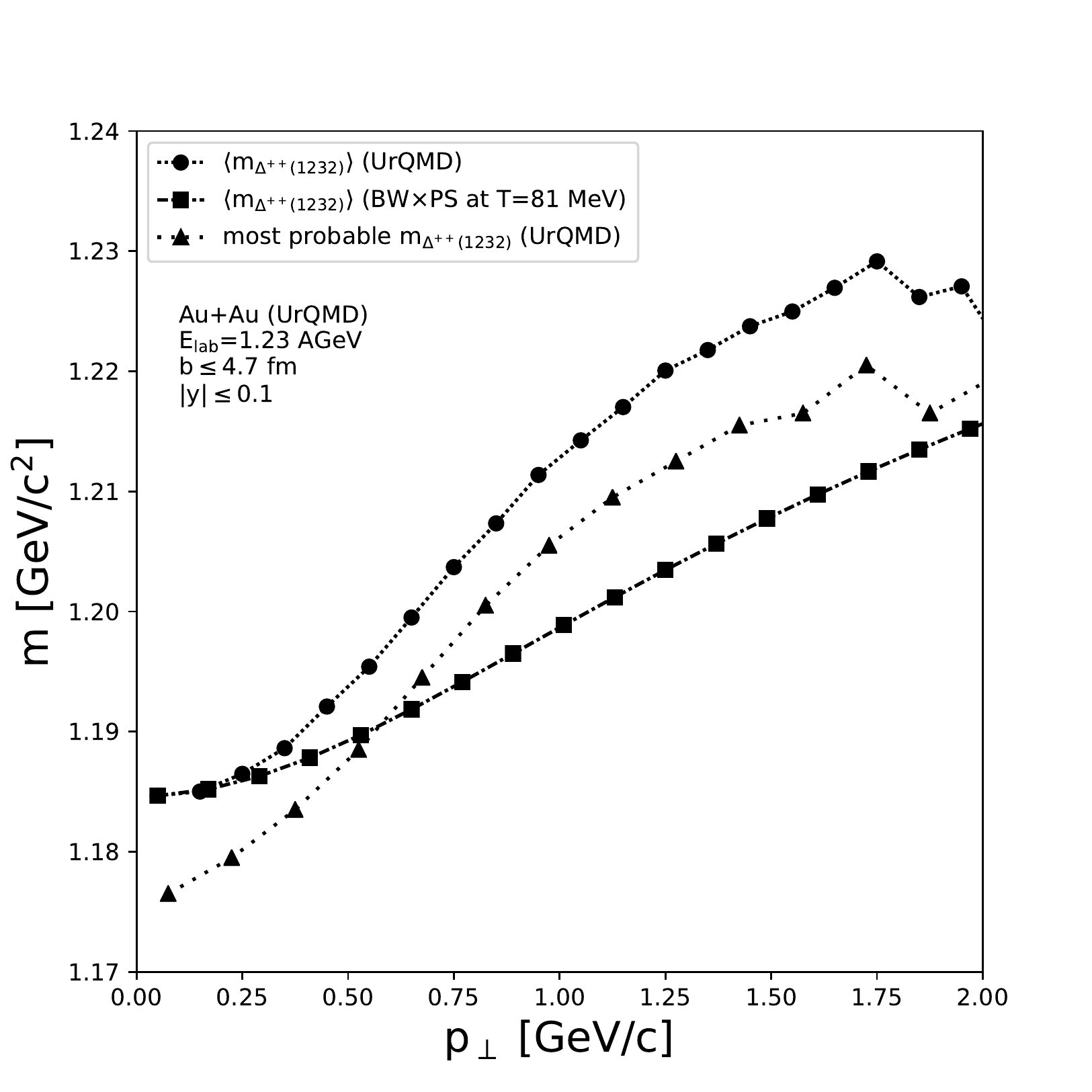}
	\caption{[Color online] Mean mass of the reconstructable $\Delta$(1232) resonances as a function of transverse momentum at midrapidity in central Au+Au reactions at 1.23 AGeV from UrQMD (line with full circles) as well as the most probable masses of the $\Delta^{++}$(1232) for several p$_{\perp}$-intervals (line with full triangles). The mean masses extracted from a simple Breit-Wigner distribution multiplied by a phase space factor for $T=81$~MeV is shown in comparison as line with full squares.
	}\label{meanm_pt}
\end{figure}

We summarize the results in Fig. \ref{meanm_pt}. Here we show the mean mass (i.e. averaged over the mass region up to 1.4~GeV/c$^2$) of reconstructable $\Delta$(1232) resonances (line with full circles) as a function of transverse momentum as well as the most probable mass of the resonances (line with full triangles) at midrapidity in central Au+Au reactions at 1.23 AGeV from UrQMD. The calculation clearly shows a decrease of the $\Delta$(1232) mass towards lower transverse momenta. In the simulation, we observe a downward shift of the mean $\Delta$ mass by 50 MeV as compared to the nominal mass. Qualitatively, this behaviour is in line with our BW times phase space argument presented above and shown for a temperature of $T=81$~MeV in comparison as a line with full squares in Fig. \ref{meanm_pt}.

\begin{figure}[t]	
	\includegraphics[width=0.5\textwidth]{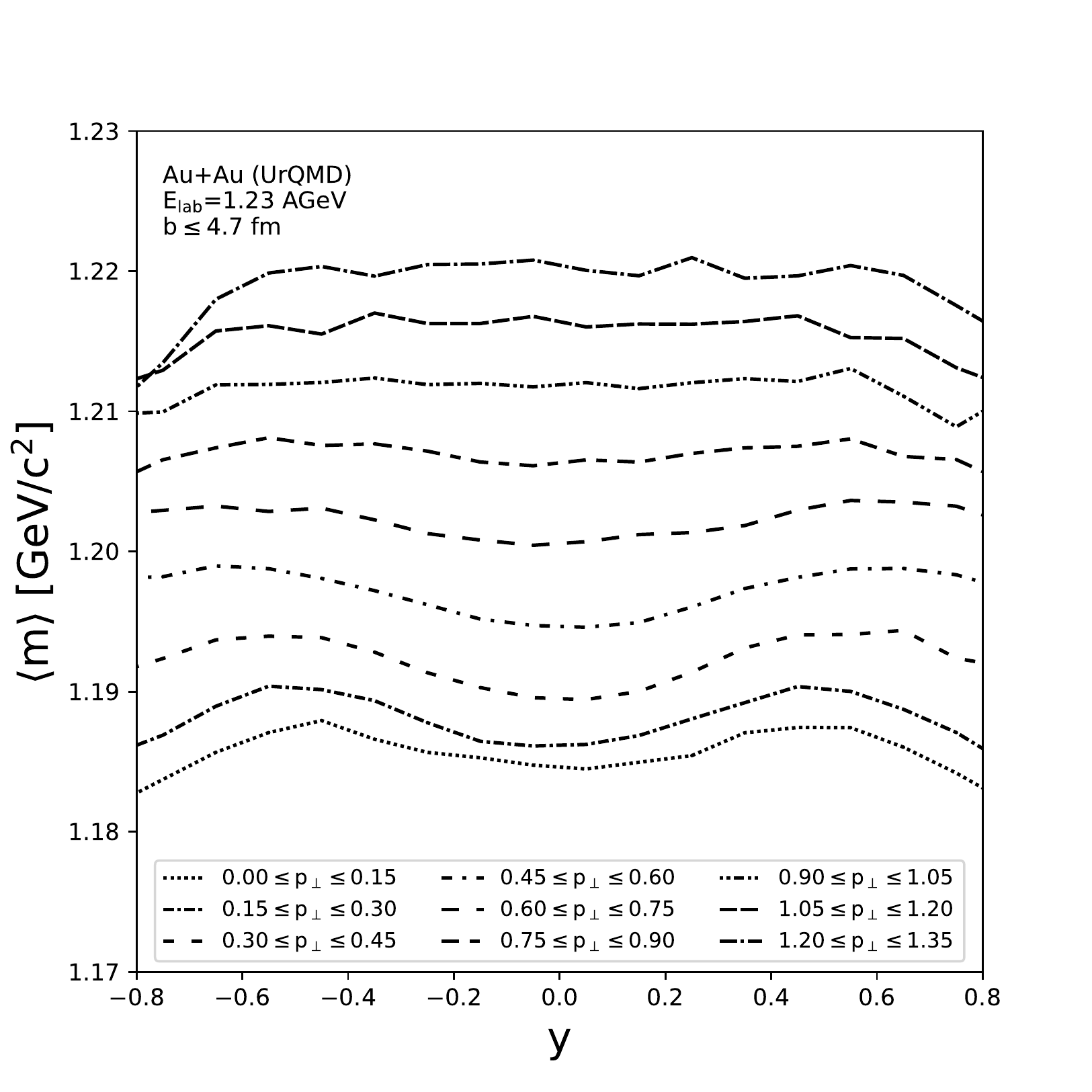}
	\caption{[Color online] Mean mass of the reconstructable $\Delta$(1232) resonances as a function of rapidity for various transverse momentum regions (from $p_{\perp}=0-0.15$ GeV to $p_{\perp}=1.2-1.35$ GeV (from bottom to top)) in central Au+Au reactions at 1.23 AGeV from UrQMD.}\label{meanm_y}
\end{figure}

Finally, we explore the rapidity dependence of the mean mass of the $\Delta$(1232) resonance in Fig. \ref{meanm_y}. From top to bottom, the analysed transverse momentum regions decrease. While for high transverse momenta the rapidity dependence is rather flat, we predict that for $p_{\perp}\rightarrow 0$ a non-trivial structure emerges. For such low transverse momenta a clear local minimum of the $\Delta$(1232) mass at midrapidity emerges. This supports the interpretations that the $\Delta$(1232) mass shift is correlated with a long evolving  $\Delta\leftrightarrow \pi+N$ cycle that is most prominent at midrapidity.




\section{Summary}
In the present paper, we predict a moderate downward mass shift of $\Delta$(1232) resonances in Au+Au reactions at 1.23 AGeV. We predict two central features: 1) The average mass (and the most probable mass) of the $\Delta$(1232) decreases with decreasing kinetic decoupling temperature and 2) the average mass (and the most probable mass) of the $\Delta$(1232) increases with increasing transverse momentum. In addition we support this finding by showing that a non-trivial rapidity dependence of the $\Delta$(1232) mass should be observed. We relate our findings to the kinetic decoupling stage of the reaction and propose to use the mass shift at low transverse momenta as thermometer of the kinetic decoupling stage. These studies are motivated by the upcoming data from the HADES collaboration for Au+Au reactions at 1.23 AGeV, which will address the production and properties of $\Delta$(1232) resonances in the channel $\Delta\rightarrow p+\pi^+$.

\section{Acknowledgments}
This work was supported by Thailand Research Fund (TRF-RGJ PHD/0185/2558), Deutscher Akademischer Austauschdienst (DAAD), HIC for FAIR and in the framework of COST Action CA15213 THOR. We thank Georgy Kornakov for an inspiring talk and discussions on this topic at the EMMI Workshop ''Probing dense baryonic matter with hadrons: Status and Perspective'' held in Darmstadt in February 2019.

\end{document}